\documentclass[preprint, onecolumn, superscriptaddress]{revtex4}
\usepackage{graphicx}
\DeclareGraphicsExtensions{. jpg, . pdf,  . mps,  .png,  .eps,  . ps,  . EPS, .gif }
\usepackage{amsmath}
\usepackage{color}
\def\be{\begin{equation}}
\def\ee{\end{equation}}

\def\bc{\begin{center}}
\def\ec{\end{center}}
\def\bea{\begin{eqnarray}}
\def\eea{\end{eqnarray}}
\newcommand{\avg}[1]{\langle{#1}\rangle}
\newcommand{\Avg}[1]{\left\langle{#1}\right\rangle}

\begin{document}
\title{Emergent Complex Network Geometry}

\author{Zhihao Wu}
\affiliation{
Beijing Key Lab of Traffic Data Analysis and Mining,  School of Computer and Information Technology,   Beijing Jiaotong University,  Beijing 100044,  China}
\author{Giulia Menichetti} 
\affiliation{ Department of Physics and Astronomy and INFN Sez. Bologna,  Bologna University,  Viale B. Pichat 6/2 40127 Bologna,  Italy, }
\author{Christoph Rahmede}
\affiliation{ Karlsruhe Institute of Technology,  Institute for Theoretical Physics,  76128 Karlsruhe,  Germany}
\author{Ginestra Bianconi\footnote{Corresponding author: ginestra.bianconi@gmail.com}}
\affiliation{School of Mathematical Sciences,  Queen Mary University of London,  E1 4NS London,  United Kingdom.}

\begin{abstract}
{\bf  Networks are  mathematical structures that are universally used to describe a large variety of complex systems  such as the brain  or the Internet.  Characterizing the geometrical properties of these networks has become increasingly relevant for routing problems, inference and data mining.
In real growing networks, topological,  structural and geometrical properties emerge spontaneously from their dynamical rules. Nevertheless we still miss 
a model in which networks develop an emergent complex geometry. Here we show that a single two parameter network model,  the  growing geometrical network,   
can  generate complex network geometries with non-trivial distribution of  curvatures,  combining exponential growth and small-world properties with finite spectral dimensionality. 
In one limit,  the non-equilibrium dynamical rules of these networks can generate scale-free networks with clustering and communities,  in another limit planar random geometries 
with non-trivial modularity. Finally we find that these properties of the geometrical growing networks are present in a large set of real networks describing biological,  social and 
technological systems.}
\end{abstract}
\maketitle

\section*{INTRODUCTION}
Recently,  in the network science community \cite{Barabasi_review,  Newman_book, Doro_book, Santo},  the interest in the geometrical characterizations of real network datasets has been growing. This problem has indeed many applications related  to routing problems in the Internet \cite{Kleinberg, Boguna_navigability, Boguna_Internet, Saniee},  data mining and community detection \cite{Mahoney_communities,  Mahoney_treelike, Vaccarino1, Vaccarino2, Munoz, Jin}.  
At the same time,  different definitions of network curvatures have been proposed by mathematicians \cite{Ollivier, Yau1, Yau2, Jost, Keller1,Keller2,Higuchi,Gromov,Knill1, Knill2},  and the characterization of the hyperbolicity of real network datasets has been gaining momentum thanks to the formulation of  network models embedded in hyperbolic planes \cite{Nechaev1, Nechaev2, Aste, Boguna_hyperbolic, Boguna_growing},  and by the definition of delta hyperbolicity of   networks by  Gromov \cite{Gromov, Mahoney_gromov, Jonck, Bary}.
This debate on geometry of networks includes also the discussion of useful metrics for spatial networks \cite{Barthelemy, Havlin_spatial}  embedded into a physical space  and its technological application including wireless networks \cite{Wireless}.
 
In the apparently unrelated field of quantum gravity,  pregeometric models,  where space is an emergent property of a network  or of a  
simplicial complex, have attracted large interest over the years \cite{Wheeler,pregeometry_review, pregeometry2,CDT1, CDT2, graphity_rg,graphity,netcosmology}. 
Whereas in the case of quantum gravity the aim is to obtain a continuous spacetime structure at large scales,  the underlying simplicial
structure from which geometry should emerge bears similarities to networks. Therefore we think that similar models 
taylored more specifically to our desired network structure (especially growing networks) could develop emergent geometrical 
properties as well.

Here our aim is to propose a pregeometric model for emergent complex network geometry,  in which the non-equilibrium dynamical rules do not take into account any embedding space,  but during its evolution the network develops a certain heterogeneous distribution of  curvatures,  a small-world topology characterized by high clustering and small average distance,  a modular structure and  a finite spectral dimension. 

In the last decades the most popular framework for describing the evolution of complex systems has been the one of growing network models \cite{Barabasi_review, Newman_book, Doro_book}. In particular growing complex networks evolving by the preferential attachment mechanism have been widely used to explain the emergence of the scale-free degree distributions  which are ubiquitous in complex networks. In this scenario,  the network grows by the addition of new nodes and these nodes are more likely to link to nodes already connected to many other nodes according to the preferential attachment rule. 
In this case the probability that a node acquires a new link is proportional to the degree of the node.
The simplest version of these models,  the Barabasi-Albert (BA) model \cite{BA},  can be modified \cite{Barabasi_review, Newman_book, Doro_book} in order to describe complex networks that also have a large clustering coefficient,  another important and ubiquitous property of complex networks that  characterizes  small-world networks \cite{WS} together with the small typical distance between the nodes. 
Moreover, it has been recently observed \cite{TriadicClosure, Redner}
that growing network models inspired by the BA model and enforcing  a high clustering coefficient, using the so called triadic closure mechanism, are able to display a non trivial community structure \cite{Ravasz, Newman}. Finally, complex social, biological and technological networks not only have high clustering but also have a structure which suggests that the networks have an hidden embedding space, describing the similarity between the nodes. For example the local structure of protein-protein interaction networks, analysed with the tools of graphlets, 
suggests that these networks have an underlying non-trivial geometry \cite{Natasa1,Natasa2}.

Another interesting approach to complex networks suggests that network models evolving in a   hyperbolic plane might model and  approximate a large variety of complex networks \cite{Boguna_hyperbolic, Boguna_growing}. In this framework nodes are embedded in a hidden metric structure of constant negative curvature that determine their evolution in such a way that nodes closer in space are more likely to be connected. 

But is it really always the case that the hidden embedding space is causing the network dynamics or might it be that this effective hidden metric space is the outcome of the network evolution?

Here we want to adopt a growing network framework in order to describe the emergence of geometry in evolving networks. We start from non-equilibrium growing dynamics independent of any hidden embedding space,  and we show that spatial properties of the network emerge spontaneously.
These networks are the skeleton of growing simplicial complexes  that are constructed by gluing together simplices of given dimension. In particular in this work we focus on simplicial complexes built by gluing together triangles and imposing that the number of triangles incident to a link cannot be larger than a fixed number $m$ that parametrizes the network dynamics.
In this way we provide evidence  that the proposed  stylized model,  including only two parameters,  can give rise to a wide variety of network geometries and can be considered a starting point for characterizing emergent space in complex networks. 
Finally we compare the properties of real complex system datasets with the structural and geometric properties of the growing geometrical model showing that despite the fact that the 
proposed model is extremely stylized,  it captures main features observed in a large variety of  datasets. 

\section*{RESULTS}

Metric spaces  satisfy the triangular inequality. Therefore in  spatial networks we must have that if a node $i$ connects two nodes (the node $j$ and the node $k$),  these two must be connected by a path of short distance. Therefore,  if we want to describe the spontaneous emergence of a discrete geometric space,  in  absence of an embedding space and a metric,  it is plausible that starting from growing  simplicial complexes should be 
an advantage. These structures are formed by gluing together complexes of dimension $d_n>1$,  i.e. fully connected networks,  or cliques,  formed by $n=d_n+1>2$ nodes,  such as triangles,  
tetrahedra etc.
For simplicity,  let us here consider growing networks constructed by addition of connected complexes of dimension $d_n=2$,  i.e. triangles.
We distinguish between two cases: the case in which a link can belong to an arbitrarily large number of triangles ( $m=\infty$),  and the case in which each link can belong at most to a 
finite number $m$ of triangles. In the case in which $m$ is finite we call the links to which we can still add at least one triangle unsaturated. All the other links we call saturated.

\begin{figure}
\includegraphics[width=8cm]{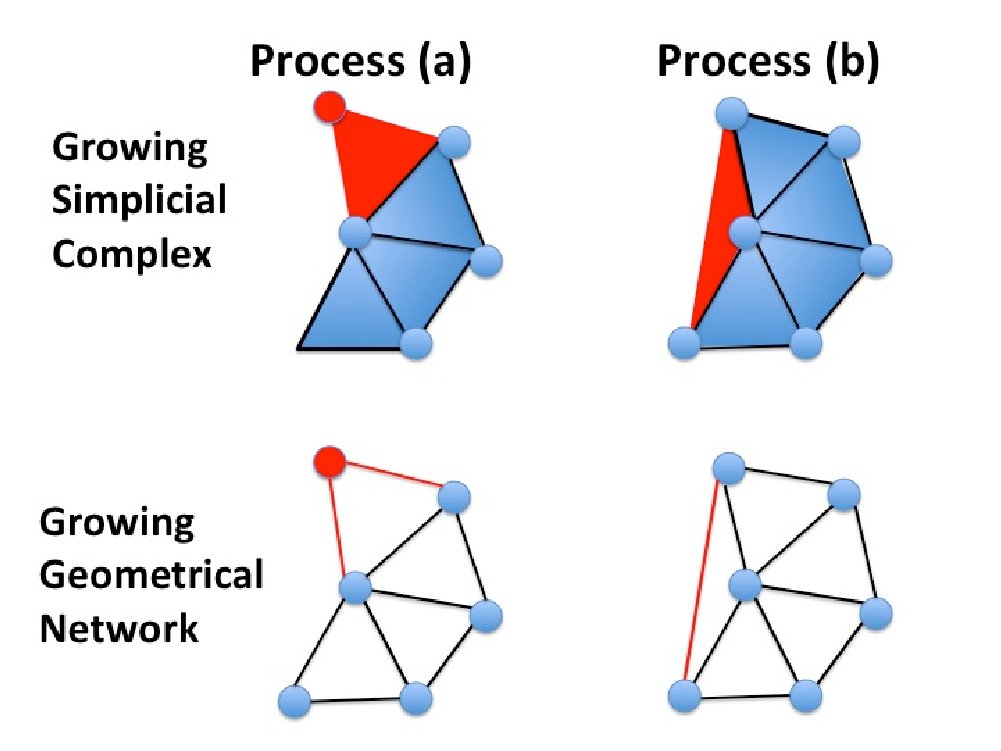}
\caption{{\bf The two dynamical rules for constructing the growing simplicial complex and the  corresponding growing geometrical network}.
In process (a) a single triangle with one new node and two new links is added to a random unsaturated link,   where by unsaturated link we indicate   a link having less than $m$ triangles incident to it. In process (b) with probability $p$ two nodes at distance two in the  simplicial complex are connected and all the possible triangles that can link these two nodes are added as long as this is allowed (no link acquires more than $m$ triangles incident to it). The growing geometrical network is just the network formed by the nodes and the links of the growing simplicial complex. In the Figure we show the case in which $m=2$.}
\label{Figure1}
\end{figure}

To be precise,  we start from a network formed by a single triangle,  a simplex of dimension $d_n=2$. At each time we perform two processes (see Figure 1).
\begin{itemize}
\item{\em Process (a)-} We add a triangle to an unsaturated link $(i, j)$ of the network linking node $i$ to node $j$. We choose this link randomly with   probability  $\Pi_{(i, j)}^{[1]}$ given by 
\bea
\Pi_{(i, j)}^{[1]}=\frac{a_{ij}\rho_{ij}}{\sum_{r, s}a_{rs}\rho_{rs}}
\label{prob}
\eea
where $a_{ij}$ is the element $(i, j)$ of the adjacency matrix ${\bf a}$ of the network,  and where the matrix element $\rho_{ij}$ is equal to one (i.e. $\rho_{ij}=1$) if the number of triangles to which  
the link $(i, j)$ belongs is less than $m$,  otherwise it is zero (i.e. $\rho_{ij}=0$).
Having chosen the link $(i, j)$ we add a  node $s$,  two links $(i, s)$ and $(j, s)$ and the new triangle linking node $i$,  node $j$ and node $s$.
\item{\em Process (b)-}	
With probability $p$ we add a single link between two nodes at hopping distance $2$,  and we add all the triangles that this link closes,  without adding more than $m$ triangles to each link. 
In order to do this,  we choose an unsaturated  link $(i, j)$ with probability $\Pi_{(i, j)}^{[1]}$ given by Eq. $(\ref{prob})$,  then we choose one random unsaturated link adjacent either to node $i$ or node $j$ as long as this link is not already part of a triangle including node $i$ and node $j$. Therefore we choose the link $(r, s)$ with probability $\Pi^{[2]}_{r, s}$ given by
\bea
\Pi^{[2]}_{r, s}=\frac{1}{\cal N}\left[{a_{is}\rho_{is}\delta_{r, i}+a_{rj}\rho_{rj}\delta_{j, s}-a_{is}\rho_{is}a_{sj}\rho_{sj}\delta_{i, r}-a_{ri}\rho_{ri}a_{rj}\rho_{rj}\delta_{j, s}}\right]
\eea
where $\delta_{x, y}$ is the Kronecker delta and ${\cal N}$ is the normalization constant.
 Let us assume without loss of generality that the chosen link $(r, s)=(r, j)$.  Then we add a link $(i, r)$ and all the triangles  passing through node $i$ and node $r$ as long as this process is allowed (i.e. if by doing so we do not add more than $m$ triangles to each link). Otherwise we do nothing. 
 \end{itemize}
 With the above algorithm (see Supplementary Information for the MATLAB code) we describe a growing simplicial complex formed by adding triangles. From this structure we can extract the corresponding network where we consider only the information about node connectivity (which node is linked to which other node).
 We call this network model the geometrical growing network.
In Figure 1 we show schematically the dynamical rules for building the growing simplicial complexes and the geometrical growing networks that describe its skeleton.

\begin{figure}
\includegraphics[width=16cm]{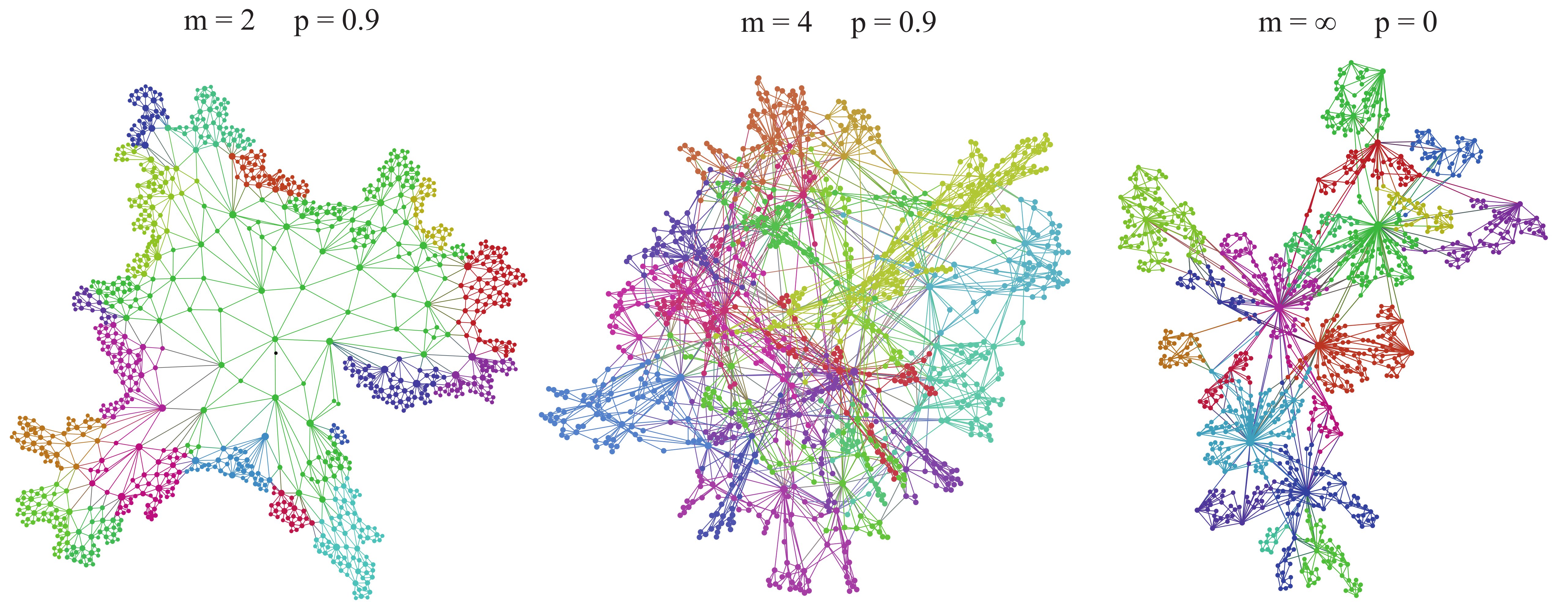}
\caption{{\bf The growing geometrical network model can generate networks with different topology and geometry.} In the case $m=2$,   $p=0.9$ a random planar geometry is formed. In the case $m=\infty$,   $p=0.9$ a scale-free network with power-law exponent $\gamma=3$ and non trivial community structure and clustering coefficient is formed. In the intermediate case $m=4,  \ p=0.9$ a network with broad degree distribution,   small-world properties and finite spectral dimension is formed. The colours here indicate division into communities found by running the Leuven algorithm \cite{Leuven}.}
\label{Figure2}
\end{figure}

Let us comment on two fundamental limits of this dynamics.
In the case $m=\infty$, $p=0$, the network is scale-free and in the class of growing networks with preferential attachment. In fact the probability that we add a link to a generic node $i$ of the network using process $(a)$
is simply proportional to the number of links connected to it, i.e. its degree $k_i$. Therefore, the mean-field equations for the degree $k_i$ of a generic node $i$ are equal to the 
equations valid for the BA model, i.e. they yield a scale-free network with power-law exponent $\gamma=3$. Actually this limit of our model  was  already discussed  in \cite{Doro_model} as a simple and major  example of scale-free network. For $m=2$, instead,   the degree distribution can be shown to be exponential (see Methods and Supplementary material for details).
The Euler characteristic $\chi$ of our simplicial complex and the corresponding network is given by   
\bea
\chi=N-L+T
\eea
where $N$ indicates the total number of nodes, $L$ the total number of links and $T$ the total number of triangles in the network.
For  $m=2$  and any value of $p$, or  for $p=0$ and any value of $m$ the networks are planar graphs since the non-planar subgraphs $K_5$ (complete graph of five nodes) and $K_{3,3}$ (complete bipartite graph formed by two sets of three nodes) are excluded from the dynamical rules (see Methods for details).  Therefore in these cases we have an  Euler characteristic $\chi=1$ (in fact here we do not count the external face).

In general the proposed growing geometric network model can generate a large variety of network geometries.
In Figure 2 we show a visualization of single instances of the growing geometrical networks in the cases $m=2$,  $p=0.9$ (random planar geometry),  $m=\infty$,  $p=0.$ (scale-free geometry),  
and $m=4$,  $p=0.9$.

\begin{figure}
\includegraphics[width=10cm]{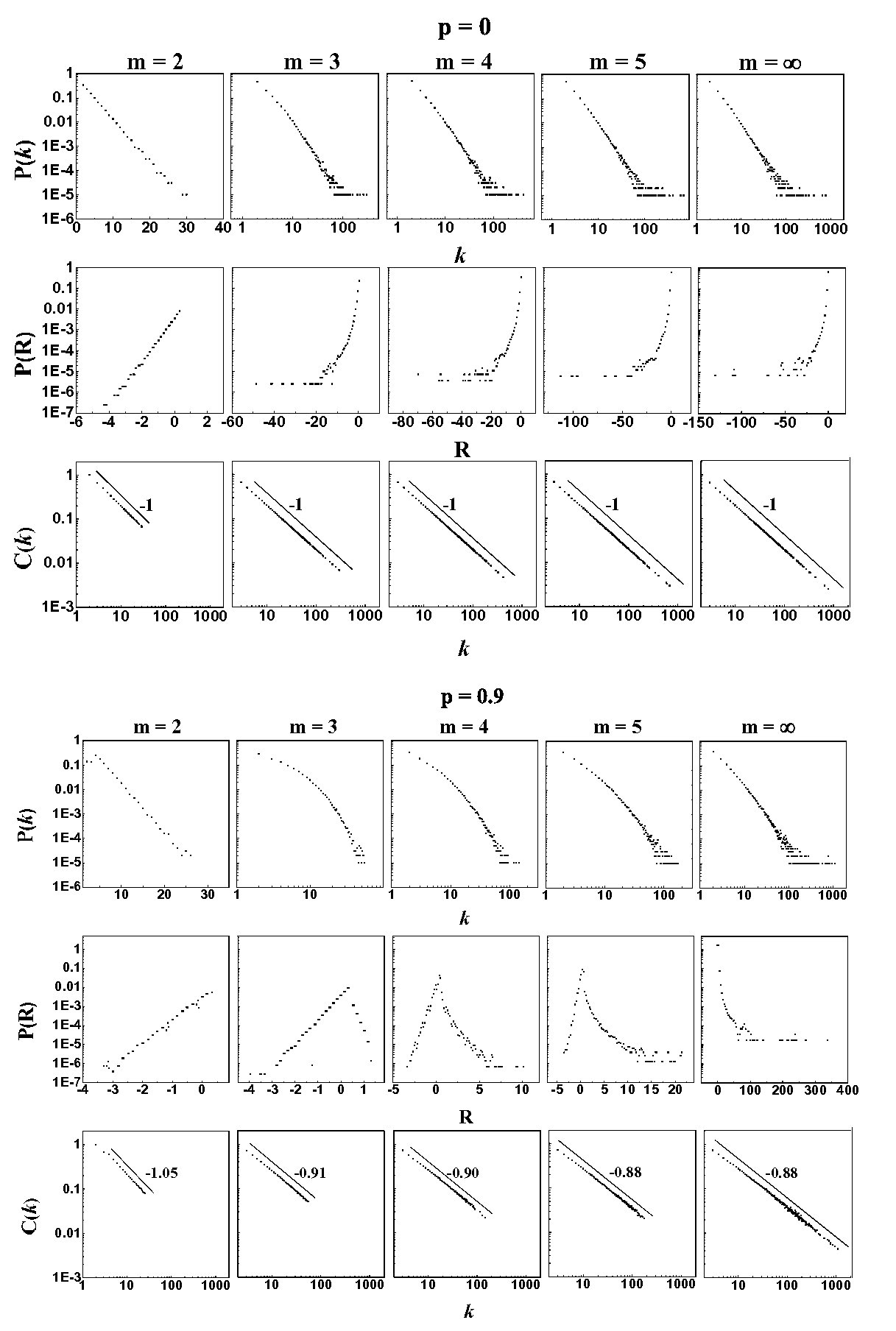}
\caption{{\bf Local properties of the growing geometrical model.}
We plot the degree distribution $P(k)$,   the distribution of  curvature $P(R)$,   and the average clustering coefficient $C(k)$ of nodes of degree $k$ for networks of sizes $N=10^5$,   parameter $p$ chosen as either $p=0$ or $p=0.9$,   and different values of $m=2,  3,  4,  5,  \infty$. The network has exponential degree distribution for $m=2$ and scale-free degree distribution for $m=\infty$. For $p>0$ and $m>2$  it shows broad degree distribution. The  networks are always hierarchical,   to the extent that $C(k)\simeq k^{-\alpha}$ with $\alpha$ shown in the figure. The distribution of  curvature $R$ is exponential for $m=2$ and scale-free for $m=\infty$. For $\alpha<1$ the curvature has a positive tail.  }
\label{Figure3}
\end{figure}
The growing geometrical network model has just two parameters $m$ and $p$.
The role of the parameter $m$ is to fix the maximal number of triangles incident on each link. 
The role of the parameter $p$ is to allow for a non-trivial K-core structure of the network. 
In fact,  if $p=0$ the network can be completely pruned if we remove nodes of degree $k_i=2$ recursively,  similarly to what happens in the BA model,  while for $p>0$ the geometrical 
growing network has a non-trivial $K$-core. Moreover the process $(b)$ can be used to ``freeze" some region of the network. 
In order to see this,   let us consider the role of the process $(b)$ occurring with probability  $p$ in the case of a network with $m=2$.
Then for $p=0$,  each node will increase its connectivity indefinitely with time having always exactly two unsaturated links attached to it. On the contrary,  if $p>0$ there is a small probability that some nodes will have all adjacent links saturated,  and a degree that is frozen  and does not grow any more. A typical network of this type is shown for  $m=2, p=0.9$ in Figure 2 where one can clearly distinguish between an active boundary of the network where still many triangles can be linked and a frozen bulk region of the network.

\begin{figure}
\includegraphics[width=12cm]{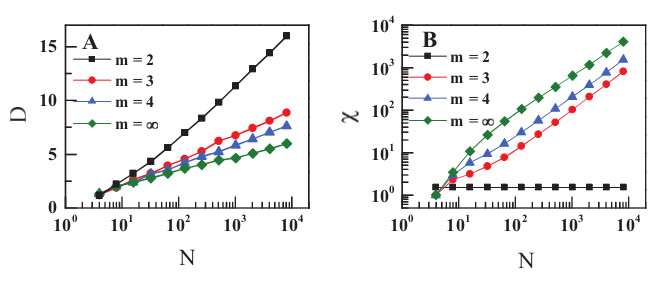}
\caption{{\bf Maximum distance $D$ from the initial triangle and Euler characteristic $\chi$ as a function of the network size $N$.} 
The  geometrical network model is growing exponentially,   with $D(N)\propto \log(N)$. Here we show the data  $m=2,  3,  4,  5,  \infty$ and $p=0.9$ (panel A). The Euler characteristic $\chi$ is given by $\chi=1$ for $m=2$ and $p=0$ and grows linearly with $N$ for the other values of the parameters of the model (panel B).
}
\label{Figure5}
\end{figure}
The geometrical growing networks have highly heterogeneous structure reflected in their local properties.
For example,   the degree distribution is scale-free for $m=\infty$ and exponential for $m=2$ for any value of $p$. Moreover  for finite values of $m>2$ the degree distribution can develop a tail that is broader for 
increasing values of $m$ (see Figure 3).  
Furthermore,  in  Figure 3 we plot the average clustering coefficient $C(k)$ of nodes of degree $k$ showing that the geometrical growing networks are hierarchical \cite{Ravasz}, they have a clustering coefficient $C(k)\propto k^{-\alpha}$ with values of $\alpha$ that are typically  $\alpha\leq 1$.

Another important and geometrical local property is the   curvature, defined on each node of the network.
For  either $m=2$ and any value of $p$ or for $p=0$ and any value of $m$, the generated graph is a planar network of which all faces are triangles.
Therefore we consider the curvature  $R_i$ \cite{Keller1,Keller2,Gromov,Higuchi} given by
\bea
R_i=1-\frac{k_i}{2}+\frac{t_i}{3},
\label{lK}
\eea
where $k_i$ is the degree of node $i$, and $t_i$ is the number of triangles passing through node $i$.

We observe that the definition of the curvature satisfies the Gauss-Bonnet theorem
\bea
\chi=\sum_{i=1}^N R_i.
\eea
For a planar network, for bulk nodes which have $k_i=t_i$ the curvature reduces to 
\bea
R_i=1-\frac{k_i}{6}
\eea
and for nodes at the boundary for which $k_i=t_i+1$, it reduces to
\bea
R_i=\frac{4-k_i}{6}.
\label{rp}
\eea 
Note  that the expression in Eq.~$(\ref{rp})$ is also valid for $m>2$ as long as $p=0$. In fact for these networks only  process $(a)$ takes place and it is easy to show that $k_i=t_i+1$. 
This simple relation between the curvature $R_i$ and the degree $k_i$ allows to characterize the distribution of curvatures in the network easily.
The curvature is intuitively related to the degree of the node. As all triangles are isosceles, a bulk node with degree six has zero curvature. In fact the sum of the angles of the  triangles incident to the node is $2\pi$. Otherwise the sum is smaller or larger than $2\pi$ resulting in positive or negative curvature respectively. The argument works similarly for the nodes at the boundary.

\begin{figure}
\includegraphics[width=16cm]{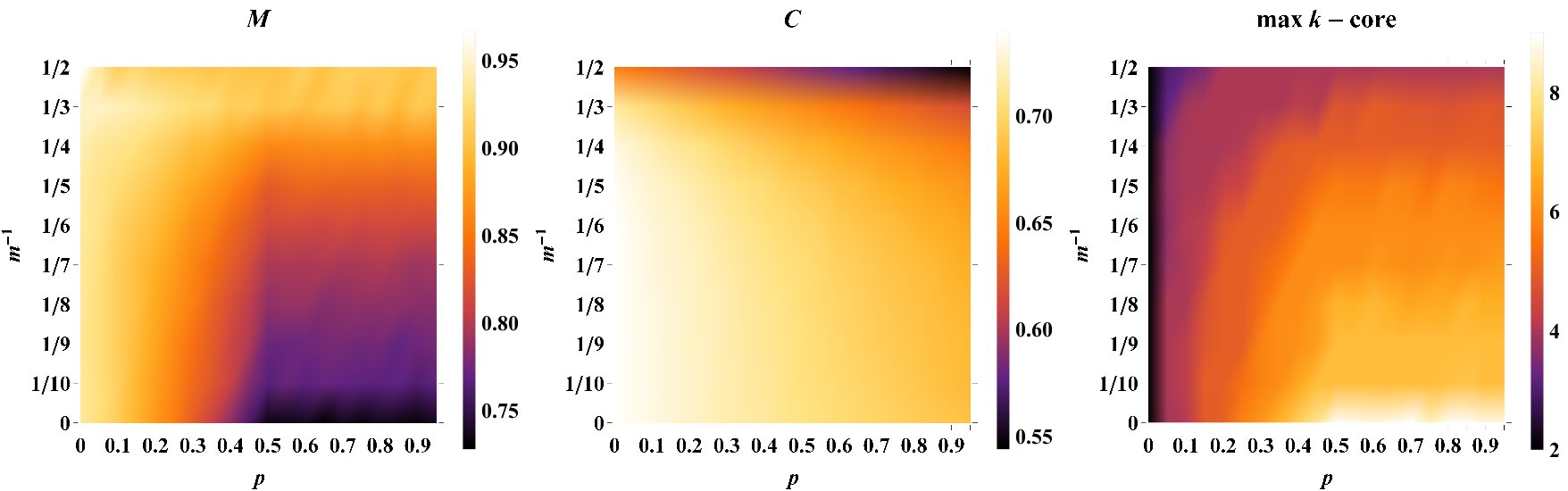}
\caption{{\bf Modularity and clustering of the growing geometrical model.} 
The modularity $M$ calculated using the Leuven algorithm \cite{Leuven} on $10$ realisations of the growing geometrical network of size $N=10^5$ is reported as a function of the parameters $m$ and $p$ of the model. Similarly the  average local clustering coefficient $C$ calculated over $10$ realisations of the growing geometrical networks of size $N=10^5$ is reported as a function of the parameters $m$ and $p$. The value of $K$ of the maximal $K$-core is shown for a network of $N=10^4$ nodes as a function of $m$ and $p$. These results show that the growing geometrical  networks have finite average clustering coefficient together with non-trivial community and $K$-core structure on all the range of parameters $m$ and $p$.}
\label{Figure4}
\end{figure}

\begin{figure}
\includegraphics[width=12cm]{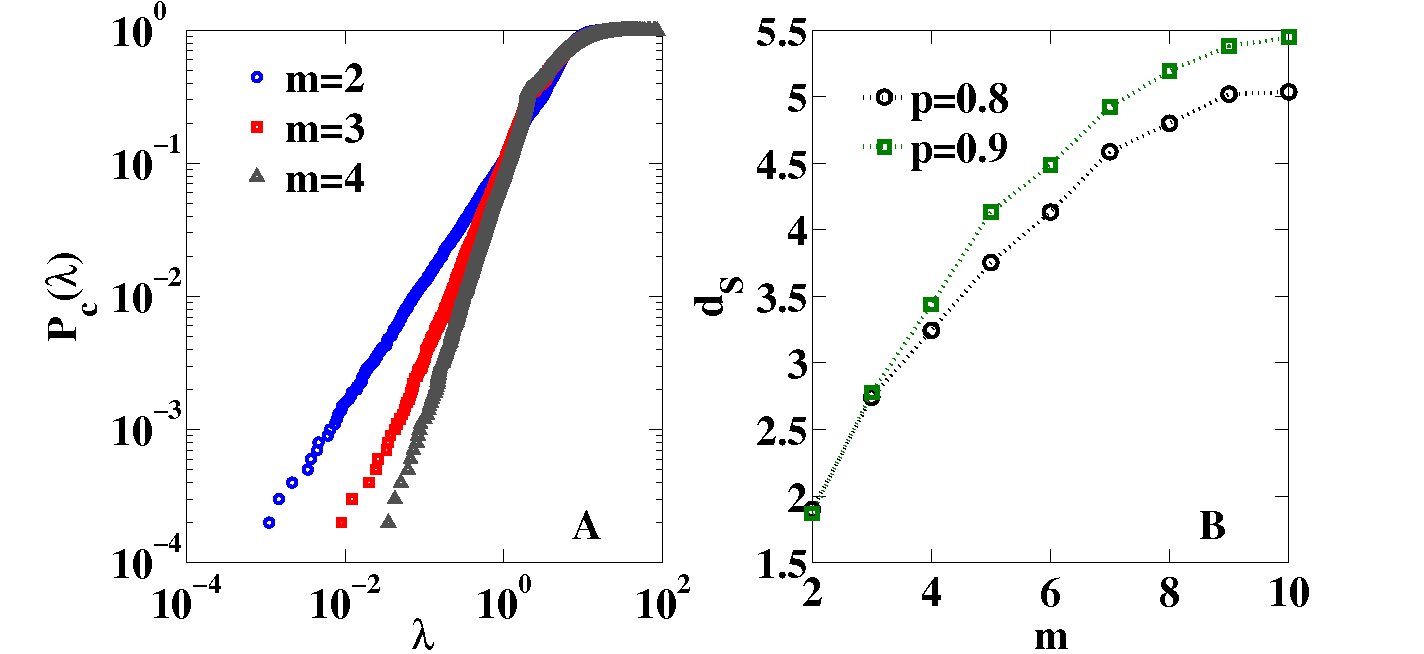}
\caption{{\bf The spectral dimension of the geometrical growing networks.} 
Asymptotically in time,   the geometrical growing networks have a finite spectral dimension.
Here we show typical plots of the spectral density of networks with $N=10^4$ nodes, $p=0.9$ and $m=2,  3,  4$ (panel A).
In panel B we plot the fitted spectral dimension for $N=10^4$ averaged over $40$ network realizations for $p=0.8,0.9$.
}
\label{Figure6}
\end{figure}

For $m>2$ and $p>0$ the networks are not planar anymore, and the definition of  curvature is debated \cite{Yau1,Yau2,Jost,Ollivier}.
Here we decided to continue to use the definition given by Eq. $(\ref{lK})$. 
This is equivalent to the definition of curvature by Oliver Knill \cite{Knill1,Knill2}, 
in which the curvature $R_i$ at a node $i$ is defined as 
\bea
R_i=\sum_{n=1}^{N} (-1)^{n+1} \frac{V_i^n}{n}
\label{Ru}
\eea
where $V_i^{(n)}$ are the number of simplices of $n$ nodes and dimension $d_n=n-1$ to which node $i$ belongs. In fact the definition  of curvature given by Eq. $(\ref{lK})$  is equivalent to the definition given by Eq.$(\ref{Ru})$ if we truncate the sum in Eq.$(\ref{Ru})$ to simplices
of dimension $d_n\le 2$, i.e. we consider only nodes, links and triangles since these are the original simplices building our network. 
\\
For $p=0$ the curvature distribution is dominated by a negative unbounded tail that is exponential in the case $m=2$ and power-law in the case $m=\infty$. 
In particular while the average curvature is $\avg{R}=0$ for $p=0$ and any value of $m$, in the limit $N\to \infty$ the fluctuations around this average are finite  (i.e. $\avg{R^2}<\infty$)
for $m=2$, and infinite (i.e. $\avg{R^2}=\infty$) for $m=\infty$. We note  here that   in the BA model the clustering coefficient $C_i$ of any node $i$  vanishes in the large network limit, therefore the  curvature $R_i\simeq 1-k_i/2$, and the curvature distribution has a power-law negative tail and diverging $\avg{R^2}$ in the large network limit, similarly to the case $m=\infty$ and $p=0$ of the present model.

For a general value of $p$,  we can assume that the average clustering $C(k)$ of nodes of degree $k$,  scales as 
$C(k)\simeq k^{-\alpha}$.
Then the average number of triangles $t(k)$ of nodes of degree $k$ scales as
$t(k)=k(k+1)C(k)/2\simeq k^{2-\alpha}$.  Therefore,  for large $k$ and as long as $\alpha<1$  the average curvature  of nodes of degree $k$ $ R(k)=\Avg{R_i}_{k_i=k}$, is dominated by the contribution 
of triangles and scales like $R(k)\simeq k^{2-\alpha}$ with a positive tail for large values of $k$. This allows us to distinguish the phase diagram in two different 
regions according to the value of the exponent $\alpha$: the case $\alpha<1$ in which the curvature has a positive tail, and the case $\alpha=1$ in which the curvature can have  
a negative tail.
\\
We make here two main observations. First of all,  with the definition of the curvature given by Eq. $(\ref{lK})$,  our network model has heterogeneous distribution of  curvatures. 
Therefore here we are characterizing highly heterogeneous geometries and the geometrical growing network does not have a constant curvature.
This is one of the main differences of the present model compared to  network models embedded in the hyperbolic plane \cite{Boguna_hyperbolic, Boguna_growing}.
In particular all the networks with $m=2$ or $p=0$ have $\chi=1$ and therefore the average curvature  is zero in the thermodynamical limit,  but they have a curvature distribution with 
an unbounded negative tail that can be either exponential for $m=2$ (i.e. $\avg{R^2}<\infty$) 
or scale-free as for the case $m=\infty$ 
(i.e. $\avg{R^2}=\infty$).
\\
We illustrate this in Figure 3 where we plot the distribution $P(R)$ of  curvatures for different specific models of growing geometrical networks for $p=0$ and $p=0.9$ for different 
values of $m$. 
 We show that for $p=0$ the negative tail can be either exponential or scale-free. For $p=0.9$ we have for $m=2$ a negative exponential tail and for $m=\infty$ a positive 
 scale-free tail of the curvature distribution  consistent with a value of the exponent $\alpha<1$ and a power-law degree distribution.
 \\
Our second observation is that the case $m=2$ and $p=0$ is  significantly different from the case $m>2$ and $p>0$. In fact for $m=2$ and for $p=0$ 
the Euler characteristic of the network is $\chi=1$ and never increases in time (see Methods for details),  while for the case $m>2$,  $p>0$ we expect $\chi/N$ to go to a 
finite limit as $N$ goes to infinity.
In Figure 4 the numerical results of the Euler characteristic $\chi$ as a function of the network size $N$ shows that,  for $m>2$ and $p\neq 0$,  $\chi$ grows linearly with $N$.
The quantity $\lim_{N\to \infty}\chi/N$ gives the average curvature in the network and is therefore zero for $m=2$ and $p=0$.

The generated topologies are small-world. In fact they combine high clustering coefficient with a typical distance between the nodes increasing only logarithmically with the network size. 
The exponential growth of the network is to be expected by the observation that in these networks we always have that the total number of links as well 
as the number of unsaturated links scale linearly with time. This corresponds to a physical situation in which the ``volume" (total number of links) is proportional to the ``surface" (number of unsaturated links). Therefore we should expect that the typical distance of the nodes in the network should grow logarithmically with the network size $N$.
In order to check this, 
in Figure 4 we give $D$, 
the average distance of the nodes from the initial triangle over the different network realisations
as a function of the network size $N$. From this figure it is clear that asymptotically in time $D\propto \log N$,  independently of the value of $p$ and $m$.

The effects of randomness and emergent locality in these networks are reflected by their cluster structure,  revealed by the lower bound on their maximal modularity measured by 
running efficient community detection algorithms \cite{Leuven} (Figure 5). Moreover also their clustering coefficient provides evidence for their emergent locality (Figure 5). 
Finally we observe that for $p>0$ the network develops also a non-trivial K-core structure. In order to show this in Figure 5 we also plot the value of  $K$ corresponding to the maximal $K$-core of the network.
As we already mentioned,  for $p=0$ we have $K=2$ and the network can be completely pruned by removing the triangles recursively. For $p>0$ instead,  the maximal $K$-core can have a much 
larger value of $K$,  as shown in Figure 5 for a network of $N=10^4$ nodes.
 
Therefore  these structures are different from the small world model to the extent that they are always characterised by a non-trivial community and $K$-core structure.

The geometrical growing network is growing exponentially,  so the Hausdorff dimension is infinite. Nevertheless,  these networks develop a finite spectral dimension $d_S$ as 
clearly shown in Figure 6,  for $m=2, 3, 4$  and $p=0.9$. 
We have checked that also for other values of $p$ the spectral dimension remains finite. This is a clear indication that these networks have non-trivial diffusion properties. 

The geometrical growing network model is therefore a very stylized model with interesting limiting behaviour,  in which geometrical local and global parameters can emerge spontaneously from the non-equilibrium dynamics.
Moreover here we compare the properties of the geometric growing network with the properties of a variety of real datasets.
In particular we have considered network datasets coming from biological,  social,  and technological systems and we have analysed their properties. In Table 1 we  show that in several cases 
large modularity,  large clustering,  small average distance and non-trivial maximal $K$-core structure emerge. Moreover,  in these datasets a non-trivial distribution of curvature 
(defined as in Eq.~(\ref{lK})) is present,  showing either negative or positive tail (see Figure 7). Finally the Laplacian spectrum of these networks also displays a power-law tail 
from which an effective finite spectral dimension can be calculated (see  Table 1 and Supplementary Information for details). 
This shows that the geometrical growing network models have many properties in common with real datasets,  describing biological,  social,  and technological
systems,  and should therefore be used 
and modified to model several real network datasets.

\section*{DISCUSSION}

In conclusion,  this paper shows that growing simplicial complexes and the corresponding growing geometrical networks are  characterized by the spontaneous emergence of locality and 
spatial properties. In fact small-world properties,  non-trivial community structure,  and even finite spectral dimensions are emerging in these networks despite the fact that   
their dynamical rules do not depend on any embedding space.  These growing networks are determined by non-equilibrium stochastic dynamics and provide evidence that it is possible to generate random complex self-organized geometries by simple stochastic rules.  

An open question in this context is to determine the underlying metric for these networks. In particular we believe that the investigation of the hyperbolic character of the models with $m=2$ and $p=0$ (that have zero average curvature but a negative third moment of the distribution of curvature) should  be extremely interesting to shed new light on ``random geometries" in which the curvature can 
have finite or infinite deviations from its average. A full description of their structure using tools of geometric group theory could be envisaged to solve this problem. This analysis could be facilitated also by the study of the dual network in which each triangle is a node of maximal degree $3m$. In fact each edge of the triangle is at most 
incident to other $m$ triangles in the geometrical growing network. 

Furthermore we mention that the model can be generalized in two main directions. On the one hand
the model can be extended by considering geometrical growing networks built by gluing together simplices of higher dimension.  On the other hand,  one can explore methods to generate networks that have a finite Hausdorff dimension,  i.e. that they have a typical distance between the nodes scaling like a power of the total number of nodes in the network. Another interesting direction of further theoretical investigation is to consider the equilibrium models of networks (ensembles of networks) 
in which a constraint on the total number of triangles incident to a link is imposed,  similarly to recent works that have considered ensembles with given degree correlations and average clustering coefficient $C(k)$ of nodes of degree $k$ \cite{Bsimon}.

 Finally the geometrical growing network  is a very stylized model and includes the essential ingredients for describing the emergence of locality of the interactions in complex networks and can be used in  a variety of fields in which networks and discrete spaces 
 are important,  including complex networks with clustering such as biological,  social,  and technological networks.

\begin{table}
\begin{center}
\begin{tabular}{|c|c|c|c|c|c|c|c|}
\hline
Datasets & $N$ & $L$ &$\avg{\ell}$&  $C$&$M$&$K$&$d_S$\nonumber \\
\hline
1L8W (protein)& 294&1608&5.09&0.52&0.643&7&1.95\nonumber\\
1PHP (protein)&219&1095&4.31&0.54&0.638&6&2.02\nonumber \\
1AOP chain A (protein)&265&1363&4.31&0.53&0.644&7&2.01\nonumber\\
1AOP chain B (protein)&390&2100&4.94&0.54&0.685&7&2.03\nonumber\\
Brain-(coactivation) \cite{Sporns1}&638&18625&2.21&0.384&0.426&46&4.25\nonumber \\
Internet \cite{internet}&22963&48436&3.8&0.35&0.652&25&5.083\nonumber \\
Power-grid\cite{WS} &4941&6594&19&0.11&0.933&5&2.01\nonumber\\
Add Health (school61)\cite{AddHealth} &1743& 4419&6&0.22&0.741&6&2.97\nonumber \\
\hline
\end{tabular}
\end{center}
\caption{{\bf Table showing the structural properties of a variety of real datasets.} $N$ indicates the total number of nodes,   $L$ the total number of links,   $\avg{\ell}$ the average shortest distance between the nodes,   
$C$ the average local clustering coefficient,   
$M$ the modularity found by the Leuven algorithm\cite{Leuven}, $K$ the maximal $K$-core, and $d_S$ the spectral dimension of the networks. 
 The average shortest distance $\avg{\ell}$ can be checked to be of the same order of magnitude as $L_{rand}=\log(N)/\log\avg{k}$ 
which is the average shortest distance in a random network with the same density of links as the real dataset.
The average local clustering coefficient $C$ can be checked to be much larger than  $C_{rand}=\avg{k}/N$ indicating the average clustering coefficient of a random network with the same density of links as the real dataset.  For the implications of the finite spectral dimension of proteins on their stability  see \cite{Burioni1}.
The references indicate the source of the data (for the four contact maps of the considered proteins, extracted from \cite{proteins}  see Supplementary Information for details).}
\end{table}

\begin{figure}
\includegraphics[width=12cm]{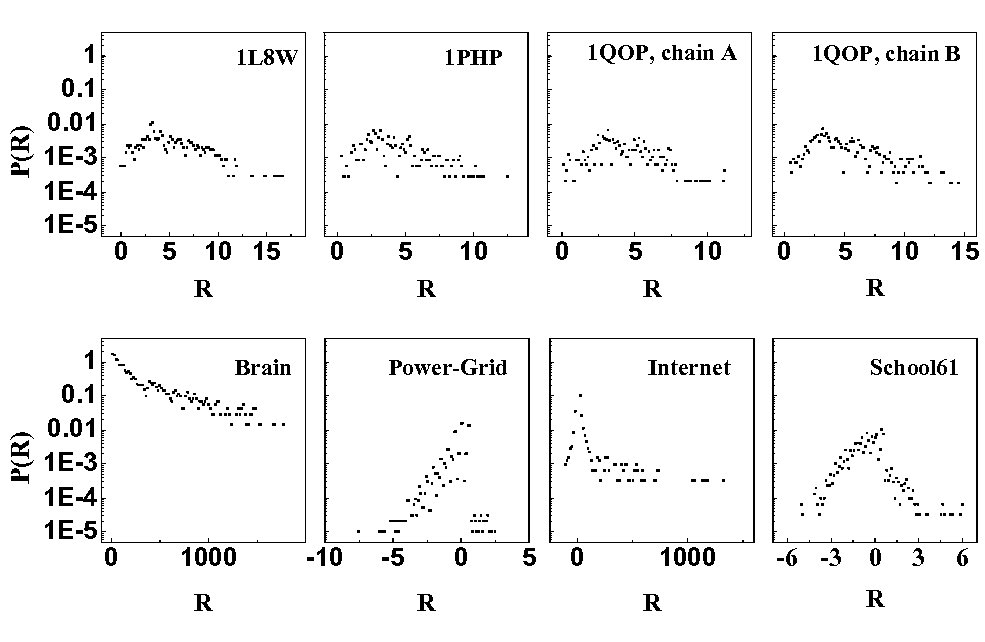}
\caption{{\bf Curvature distribution in real datasets.} 
We plot the distribution $P(R)$ in a a variety of datasets with additional structural and local properties  shown in Table 1.}
\label{Figure7}
\end{figure}

\section*{METHODS}
\subsection*{Degree distribution of $m=\infty$ and $p=0$-}
In the case $m=\infty$ and $p=0$ the geometrical growing network model is reduced to the model proposed in \cite{Doro_model}. Here we show the derivation of the scale-free distribution in this case for completeness.
In the geometrical growing network with $m=\infty$ and $p=0$ at each time a random link is chosen and a new node attaches two links to the two ends of it. Therefore the probability that at time $t$ a new link is attached to a given node of degree $k$ is given by $\frac{k}{2t}$.
Using this result we can easily write the master equation for the  number of nodes $N(k, t)$ of degree $k$ at time $t$,
\bea
N(k, t+1)=N(k, t)+\frac{k-1}{2t}N(k-1, t)[1-\delta_{k, 2}]-\frac{k}{2t}N(k, t)+\delta_{k, 2}\ .
\label{uno}
\eea
Since the network is growing,  asymptotically in time the number of nodes of degree $k$ will be proportional to the degree distribution $P(k)$,  $N(k, t)\simeq tP(k)$, where the total number of nodes in the network is $N=t+1\simeq t$. Therefore,  substituting this scaling in Eq.~$(\ref{uno})$ we get 
\bea
(2+k)P(k)=(k-1)P(k-1)
\eea
for every $k>2$,  while $P(2)=1/2$ yielding the solution 
\bea
P(k)=\frac{12}{(k+2)(k+1)k}
\eea
for $k\geq 2$,  which is equal to the degree distribution of the BA model with minimal degree equal to $2$,  i.e. scale-free with power-law exponent $\gamma=3$.
Here we observe that the curvature of the nodes is in this case $R=1-k/4$,  therefore $P(R)$ has a power-law negative tail,  i.e. $P(R)\simeq |R|^{-3}$ for $R<0$ and $|R|\gg1$. Moreover we have   $\avg{R}=0$ (consistent with $\chi=1$) but $\avg{R^2}$ is diverging with the network size $N$.
 
\subsection*{Degree distribution of $m=2$ for  $p=0$-}
The degree distribution for $m=2$ is exponential for any value of $p$. Here we discuss the simple case $p=0$ leaving the treatment of the case $p>0$ to the Supplementary Information.
For $p=0$ every node has exactly two unsaturated links. The total number of unsaturated links  is $L=1+t\simeq t$ at large time $t$. Therefore the average number of links that a node gains at time $t$  by process $(a)$ is given by 
$2/t$ for $t\gg 1$.
The master equations for the average number of nodes $N(k, t)$ that have degree $k$ at time $t$ are given by 
\bea
N(k, t+1)=N(k, t)+\frac{2}{t}N(k-1, t)-\frac{2}{t}N(k, t)+\delta_{k, 2}.
\eea
In the large time limit,  in which $N(k, t)\simeq tP(k)$, the degree distribution $P(k)$ is given by 
\bea
P(k)=\frac{1}{2}\left(\frac{2}{3}\right)^{k-1}
\eea
for $k\geq 2$.
The curvature $R=1-k/4$ is therefore in average $\avg{R}=0$ in the limit $t\to \infty$ with finite second moment $\avg{R^2}$.

\subsection*{Euler characteristic $\chi$ of geometrical growing network with either $m=2$ or $p=0$-}
The Euler characteristic of the geometrical growing networks with $p=0$ is $\chi=1$ at every time.
In fact we start from a single triangle,  therefore at $t=0$ we have $\chi=1$. At each time step we attach a new triangle to a given unsaturated link,  therefore we add one new node,  two new links,  and one new triangle,  so that 
$\Delta \chi=\Delta N-\Delta L+\Delta T=0$. Hence $\chi=1$ for every network size.
For $m=2$ also the process $(b)$ does not increase the Euler characteristic. In fact in this case when the process $(b)$ occurs,  and $m=2$,  we add only one new link and one new triangle,  therefore $\Delta \chi=0$ also for this process.
Instead in the case $m>2$ and $p>0$,  process $(b)$ always adds a single  link but the number of triangles that close is in average greater than one,  therefore the Euler characteristic  $\chi$ grows linearly with the network size $N$.

\subsection*{Definition of Modularity $M$-}

The modularity $M$ is a measure to evaluate the significance of the community structure of a network. 
It is defined \cite{Newman} as 
\bea
M=\frac{1}{2L}\sum_{ij}\left(a_{ij}-\frac{k_ik_j}{2L}\right)\delta(q_i, q_j) \ .
\eea
Here,  ${\bf a}$ denotes the adjacency matrix of the network,  $L$ the  total number of links,  and $\{q_i\}$,  where $q_i=1, 2\ldots Q$,  indicates to which community the node $i$ belongs.
Finding the network partition that optimizes modularity is a NP hard problem. Therefore different greedy algorithms have been proposed to find the community structure such as the   Leuven method \cite{Leuven} that we have used in this study. The modularity found in this way is a lower bound on the maximal modularity of the network. 
\subsection*{Definition of the Clustering coefficient-}
The clustering coefficient is given by  the probability that two nodes,  both connected to a common node,  are also connected. In the context of social networks,  it describes the probability that a friend of a friend is also your friend. 
The local clustering coefficient $C_i$ of node $i$ has been defined as the probability that two neighbours of the  node $i$  are neighbours of each other, 
\bea
C_i=\frac{t_i}{k_i(k_i-1)/2}\ , 
\eea
where $t_i$ is the number of triangles passing through node $i$,  and $k_i$ is the degree of node $i$.
\subsection*{Definition of the $K$-core-}
We define the $K$-core of a network as the maximal subgraph formed by the set of nodes that have at least $K$ links connecting them to the other nodes of the $K$-core. The $K$-core of a network can be easily obtained by pruning a given network,  i.e. by removing iteratively all the nodes $i$ with degree $k_i<K$.
\subsection*{Definition of the spectral dimension of a network-}
The Laplacian matrix of the network $L$ has elements 
\bea
L_{ij}=k_i\delta_{ij}-a_{ij}.
\eea
If the density of eigenvalues $g(\lambda)$ of the Laplacian scales like 
\bea
g(\lambda)\sim \lambda^{d_S/2-1}
\eea
with $d_S>0$,  for small values of $\lambda$,  then $d_S$ is called the spectral dimension of the network.
For regular lattices in dimension $d$ we have $d_S=d$. 
Clearly,  if the spectral dimension of a network is well defined,  then  the cumulative distribution  $P_c(\lambda)$ scales like
\bea
P_{c}(\lambda)\sim \lambda^{d_S/2}
\eea
for small values of $\lambda$.

\subsection*{Real datasets}
We analysed a large variety of biological, technological and social datasets. In particular we have considered the brain network of co-activation \cite{Sporns1}, 4 protein contact maps \cite{proteins}(see Supplementary Information for details  on the  data analysis), the Internet at the Autonomous System level \cite{internet}, the US power-grid \cite{WS}, and a social network of friendship  between high-school students coming from the Add Health dataset {AddHealth}.

\section*{Acknowledgments}
 \begin{itemize}
 \item We acknowledge interesting discussions with Mari\'an Bogu\~n\'a, Oliver Knill and Sergei Nechaev.  This work has been supported by the   National Natural Science Foundation of China (61403023). Z. W. acknowledges the kind hospitality of the School of Mathematical Sciences at QMUL.
 \item
 The authors declare that they have no
competing financial interests.
\end{itemize}

\section*{Author contribution statement}
C. R.  and G.B.  designed the research, Z. W., G. M. and G. B. wrote the codes, Z. W. and G. M. prepared figures, C. R. and G. B. wrote the main manuscript text, all authors reviewed the manuscript.

\newpage

\end{document}